\magnification=\magstep1 
\font\bigbfont=cmbx10 scaled\magstep1
\font\bigifont=cmti10 scaled\magstep1
\font\bigrfont=cmr10 scaled\magstep1
\vsize = 23.5 truecm
\hsize = 15.5 truecm
\hoffset = .2truein
\baselineskip = 14 truept
\overfullrule = 0pt
\parskip = 3 truept
\def\frac#1#2{{#1\over#2}}

\nopagenumbers
%%This command suppresses the printing of page numbers.
%%You should number the pages with blue pencil in upper right corner 
%%if you send camera-ready copy.  Of course, submission by email
%%to cmt28@wuphys.wustl.edu is preferred!!
%%
%%THE FOLLOWING THREE COMMANDS LEAVE SOME SPACE AT THE TOP OF THE LEAD PAGE. 
%%(the command "\vskip 4 truecm" actually results in about 4.5 cm of empty
%%space at the top, or about 19.6%).  The publisher will probably reset the 
%%chapter heading (your title and by-line), but you should follow my 
%%19-20% prescription anyway!  In my design I am following the Les Houches 
%%lecture notes volume produced by Nova.   If you have LOTS of authors and 
%%by-lines you may want to allow a bit less space at the top (e.g. if you 
%%have 3 or more sets of authors and institutions).
\topinsert
\vskip 3.2 truecm
\endinsert
\centerline{\bigbfont MODELING NUCLEAR PROPERTIES} 
%%If your title is longer than one line, continue thus:
\vskip 8 truept
\centerline{\bigbfont WITH SUPPORT VECTOR MACHINES}
%%Don't forget to remove the % signs from the 2 preceding lines if you
%%use them to lengthen the title!
\vskip 20 truept
%%Now comes your by-line with institutional addresses.  
\centerline{\bigifont H. Li and J. W. Clark}
\vskip 8truept
\centerline{\bigrfont McDonnell Center for the Space Sciences 
and Department of Physics}
\vskip 2 truept
\centerline{\bigrfont Washington University, St. Louis, Missouri 63130, USA} 
%%In case of multiple institutions, use the following lines, iterated
%%as necessary. 
\vskip 11 truept
\centerline{\bigifont E. Mavrommatis and S.~Athanassopoulos}
\vskip 8 truept
\centerline{\bigrfont Physics Department, Division of Nuclear \&
Particle Physics}
\vskip 2 truept
\centerline{\bigrfont University of Athens, GR-15771 Athens, Greece}
\vskip 11 truept
\centerline{\bigifont K. A. Gernoth}
\vskip 8 truept
\centerline{\bigrfont Department of Physics, University of Manchester}
\vskip 2 truept
\centerline{\bigrfont Manchester M13 9PL, United Kingdom}

\vskip 1.8 truecm

\centerline{\bf 1.  INTRODUCTION}
\vskip 12 truept
Artificial neural networks and other machine-learning strategies can provide 
a valuable complement to theory-driven models of the systematics of nuclear data.
A significant effort to exploit the potential of data-driven methodologies
receives strong motivation from the current thrust toward experimental and
theoretical exploration of nuclei far from stability.  It is made possible
by the availability of a growing body excellent experimental data on 
nuclear species numbering in the thousands.  In outline, statistical models 
based on supervised learning are developed as follows.  Suppose, for
example, we wish to predict the atomic mass $M$ of a nuclear species, or
nuclide, specifying only its mass number $A$ and atomic number $Z$,
or alternatively its proton and neutron numbers $(Z,N)$.  A learning
machine has an input interface where $(Z,N)$ are fed to the device in 
coded form and an output interface where an estimate of the
mass appears for decoding.  In between there is a system or network of 
interconnected elements that acts to process the incoming information and 
produce an appropriate output.  These processing elements may resemble biological 
neurons, receiving signals from other units through weighted connections,
and displaying nonlinear response to summed input signals.  Given
a body of training data to be used as examples of the desired
mapping, in this case $(Z,N) \to M$, a suitable learning algorithm
is used to adjust the parameters of the network, e.g., the weights of 
the connections between the processing elements, so that the learning
machine (i) generates responses at the output interface that reproduce,
or closely fit, the atomic masses of the training nuclei, and (ii) serves
as a reliable predictor of the masses of test nuclei absent from the
training set.  This second requirement is a strong one -- the system
should not merely serve as a lookup table for masses of known nuclei;
it should also perform well in the much more difficult task of 
generalization.

The last two decades have seen much activity and considerable progress
in the development and application of supervised learning machines
of the type described -- which are designed to learn by example.
The most popular implementation is the multilayer feedforward
neural network (or multilayer perceptron), taught by the backpropagation
learning algorithm in one or another of its many variations [1-3].
A significant measure of success has been achieved in constructing
global models of nuclear properties based on such neural networks,
with applications to atomic masses, neutron separation 
energies, spins and parities of nuclear ground states, stability versus 
instability, branching ratios for different decay modes, and 
beta-decay lifetimes.  (For reviews, see Ref.~[4], and for recent
results on atomic-mass prediction, see Ref.~[5].)  

The support vector machine (SVM) [6-8], a principled and powerful approach
to problems in classification and nonlinear regression, came on the
scene in the 1990s.  It has become a standard tool in statistical
modeling, and for many problems it is considered the method of 
choice.  We have begun to explore the promise of SVMs for modeling
and prediction of nuclear properties.  The first results of this
effort are reported here.  

Section 2 provides an introduction to support vector machines and the
ANOVA decomposition that facilitates their effective implementation.
Section 3 summarizes the results obtained for the atomic-mass problem,
and compares the predictive performance of the SVM models with
that of multilayer backpropagation networks and state-of-the-art
``theory-thick'' models.  Additional results and comparisons for 
beta-decay halflives and for ground-state spins and parities are 
presented in Secs.~4 and 5, respectively.  Concluding remarks 
are made in Sec.~6.
\vskip 28 truept

\centerline{\bf 2.  SUPPORT VECTOR MACHINE AND ANOVA DECOMPOSITION}
\vskip 12 truept

The support vector machine (SVM), pioneered by Vapnik [6-8], may be viewed 
as an approximate realization of the goal of structural risk minimization
[9,3].  Let $({\bf x}_1,y_1),...,({\bf x}_P,y_P)$ be a set of training 
data drawn from a function $y=f({\bf x})$.  Here, ${\bf x}$ is the input 
variable, a vector of dimension $n$, while $y$ is the output variable, 
a unique real number for given ${\bf x}$.  (In the example considered
in Sec.~1, ${\bf x}$ is a vector formed from the two components $Z$ and
$N$, while $y$ is the mass $M$.)   The support vector machine is based 
on a suitable nonlinear mapping ${\bf x} \to \varphi({\bf x})$ from the 
input space to a feature space of higher dimension $m > n$.  

Applied to the task of regression, the SVM
learning strategy begins by posing an approximation $\hat y$ to the output $y$ 
as a linear combination of certain basis functions $\varphi_i({\bf x})$
in the feature space, with corresponding linear weights connecting 
the feature space to the output space.
Thus, 
$$
{\hat y} = {\hat f}({\bf x},{\bf w}) = \sum_{j=1}^m w_j \varphi_j({\bf x})\,, 
\eqno(1)
$$
where ${\bf w}$ is an $m$-dimensional vector composed of weights
$w_j$, $j=1,\ldots,m$.  (A bias term $b$ may be included in Eq.~(1)
by starting the sum at $j=0$ and introducing $w_0 \equiv b$ and
$\varphi_0({\bf x}) \equiv 1$.) To determine the image vectors $\varphi_j({\bf x})$ 
and their weights $w_j$, consider an $\epsilon$-insensitive loss function 
defined, for input ${\bf x}$, by 
%$$
%|y - {\hat f}({\bf x},{\bf w})|_\epsilon  =
%=\cases { 0 & if $| y - {\hat f}({\bf x}, {\bf w}) | < \epsilon$ \,, \cr
%                    $|y-{\hat f}({\bf x},{\bf w})| - \epsilon$ & otherwise \,, 
%\cr}
%$$
$y - {\hat f}({\bf x},{\bf w}) - \epsilon$ in case the magnitude of the error
$y - {\hat f}$ exceeds a tolerance $\epsilon$, and taken zero otherwise.
The tolerance parameter $\epsilon$ is at the disposal of the machine's 
user.  The {\it primal} optimization problem then becomes one of minimizing the overall loss (or cost function, or empirical risk), as given by the 
sum of the individual losses for all the training patterns,
$$
E_\epsilon ({\bf w}) =  \sum_{i=1}^P \left|y_i-{\hat f}({\bf x}_i,{\bf w})
\right|_\epsilon \,, \eqno(2) 
$$
subject to the inequality $\sum_{j=1}^m w_j^2 < c_0$, where $c_0$ is
a user-determined constant.

Vapnik has shown that an equivalent solution of this constrained 
optimization problem can be obtained by solving the corresponding {\it dual 
problem}, which may be stated as follows [3].
\item{1.}
Choose a kernel of the form
$$
K({\bf x},{\bf x}_i) = \sum_{j=1}^m \varphi_j({\bf x})\varphi_j({\bf x}_i) \,,
\eqno(3)
$$
symmetrical in its vector arguments and continuous in their components,
and qualifying as an inner product in some space, so as to meet the
conditions of Mercer's theorem [10,3].
\item{2.}
Given the training sample $\{ ({\bf x}_i,y_i) \}$, $i=1, \ldots, P$,
assemble the convex functional 
$$
Q(\{\alpha_i,\alpha_i'\}) = \sum_{i=1}^P y_i(\alpha_i - \alpha_i')
-\epsilon \sum_{i=1}^P (\alpha_i + \alpha_i')
- {1 \over 2} \sum_{i=1}^P \sum_{l=1}^P ( \alpha_i - \alpha_i')
(\alpha_l - \alpha_l') K({\bf x}_i, {\bf x}_l) \,. \eqno(4)
$$
\item{3.}
Maximize $Q$ subject to the constraints
$$
\sum_{i=1}^P (\alpha_i - \alpha_i') = 0\,, \qquad 0 \leq \alpha_i\,,\,\alpha_i' 
\leq C \,, \eqno(5)
$$

\noindent
where $C$ is a user-determined constant.
The optimal approximating function then takes the forms
$$
{\hat f}_{\rm opt}({\bf x},{\bf w}) = {\bf w}^T{\bf w} 
= \sum_{i=1}^{P} (\alpha_i - \alpha_i') K({\bf x},{\bf x}_i) \,, \eqno(6)
$$
where ${\bf w}^T$ the transform of the column vector ${\bf w}$.
The subset of training patterns $i$ for which $\alpha_i - \alpha_i'$
does not vanish then defines the {\it support vectors} of the machine,
corresponding to the training examples that are the most 
salient to solution of the problem.

The parameters $\epsilon$ and $C$ provide the user with control over 
the complexity of the machine, as measured by the so-called VC dimension 
[11,3], and hence over its performance in generalization. 
Careful tuning of these parameters is necessary.

Different choices for the inner-product kernel $K({\bf x},{\bf x}_i)$ yield 
different versions of the support vector machine.  The most popular are (i) the
polynomial learning machine, corresponding to 
$$
K({\bf x},{\bf x}_i) =
({\bf x}^T{\bf x}_i + 1)^p  \eqno(7)
$$
(with user-selected power $p$),
(ii) the radial-basis function (RBF) network, corresponding to 
$$
K({\bf x},{\bf x}_i) =\exp \left( - \gamma ||{\bf x} -{\bf x}_i ||^2\right) \eqno(8)
$$
(with user-selected width parameter $\gamma$), and (iii) the 
two-layer perceptron [1-3], with 
$$
K({\bf x},{\bf x}_i) =\tanh (\beta_1 {\bf x}^T{\bf x}_i + \beta_2)  \eqno(9)
$$
(freedom in setting the parameters $\beta_1$ and $\beta_2$ being
restricted by Mercer's theorem).

We are most interested in creating predictive statistical models 
capable of estimating a real-valued function $f({\bf x})$ from given
values for its independent variables comprising ${\bf x}$.  For that
reason, we have outlined the design of SVMs for solving problems 
of nonlinear regression.  However, the support vector machine was originally 
introduced to solve yes/no classification problems, and applied to problems 
in which positive and negative cases are either separable by a hyperplane in the
input space (trivial), or not (nontrivial).  For problems that are
not linearly separable in this sense, the input vectors are mapped 
nonlinearly into a higher-dimensional feature space, in which separation 
by a hyperplane becomes possible.  The principle of structural risk 
minimization then dictates that an {\it optimal} hyperplane be sought in
this space, such that the margin of separation between positive and negative 
cases is minimized.  It is known [7,8] from general learning theory that the 
error rate of a learning machine on test data (i.e., in generalization or prediction)
is bounded by the sum of two terms, namely the error rate on the training 
data and a term involving the VC dimension.  For a linearly separable 
problem treated by a SVM, the first term is zero and the second is 
minimized.  Thus, good generalization is achieved even without 
building into the model any explicit knowledge about the problem to be 
solved, beyond the raw training data.  This desirable feature is maintained 
approximately in application of SVMs to nonseparable classification 
problems and to the generically more difficult problems of regression.

The support vector machine may be broadly viewed as a kind of 
feedforward neural network, in that the inner-product kernels 
$K({\bf x},{\bf x}_i)$ provide a layer of hidden units that effect 
nonlinear processing of the inputs and provide weighted linear outputs,
which are summed by an output unit.  As seen above, the familiar structures of
radial-basis-function networks and perceptrons with one hidden layer can 
be realized as special cases by suitable choices of kernel, as specified 
above.  But a support vector machine does more: it also embodies an algorithm 
that automatically determines the number of hidden units appropriate to 
the problem at hand, whatever the choice of kernel.  This more general 
scope of the SVM approach stands in contrast to the backpropagation 
learning algorithm [1-3], which is designed especially for training 
multilayer perceptrons.

In addition to the benefits already mentioned, the support
vector machine offers other significant advantages over the 
more traditional approaches to supervised learning based
on neural networks, which involve dependence on trial
and error, rules of thumb, and heuristics.  The support
vector machine offers a generic way to control model complexity.
The curse of dimensionality is overcome by the pivotal strategy
of introducing an inner-product kernel conforming to Mercer's
theorem and solving the constrained optimization problem in its dual 
version, thereby determining the dimension of the feature space
as the number of support vectors distilled from the training set.
The procedure naturally incorporates regularization.  The
use of the $\epsilon$-insensitive cost function (2) in the 
regression application lends robustness to the machine by avoiding 
certain drawbacks of the least-square estimator employed in
the backpropagation learning algorithm (e.g., sensitivity to outliers 
and to distributions with additive noise having a long tail).  
Importantly, the SVM is guaranteed to find a global minimum of 
the error surface.  For a more detailed and systematic development 
of the properties of SVMs, the reader is directed to Haykin's excellent 
text [3], as well as the authoritative monographs of Vapnik [7,8].

Our investigations of the potential of support vector machines for
the design of global statistical models of nuclear properties make use 
of the RBF kernel (8), as well as a simplified version of what is 
called ANOVA decomposition [12].  ANalysis Of VAriance (ANOVA) is a 
scheme for imposing a structure on multi-dimensional kernels that are 
generated from one-dimensional kernels, in a way that gives better control 
over the capacity of the machine (as measured by the VC dimension).  
An ANOVA kernel we have found to be well suited to the regression
problem posed by the nuclear (atomic) mass data is rooted in
the RBF kernel and has the form
$$
K({\bf x},{\bf x}_i) =  \left(\sum_{l=1}^n \exp\left[-\gamma\left(
  x^{(l)} - x_i^{(l)} \right)^2     \right]             \right)^d   \,,
$$
where the user-selected parameter $\gamma$ can take any positive
value and the power $d$ is usually an integer.  We 
shall call this the ANOVA kernel.
\vskip 28 truept
\centerline{\bf 3.  SVM MODELS OF NUCLEAR MASS SYSTEMATICS}
\vskip 12 truept

SVM regression models have been trained to predict $(\Delta M)c^2$
in MeV, where $\Delta M$ is the mass excess (or mass defect) 
defined by the difference $M - A$ between the atomic mass $M$,
measured in amu, and the mass number $A$ of the nuclide in question.  
In our initial study, we focus on a database given by the union 
${\rm O}\, \oplus {\rm N}\, \oplus {\rm NB}$ of three data sets.  The first
consists of the set of 1323 ``old'' (O) experimental mass assignments 
which the 1981 semi-empirical droplet-model mass formula of M\"oller 
and Nix [13] was intended to reproduce.  The second is a set of 
351 ``new'' (N) experimental mass assignments for nuclei that lie 
mostly beyond the edges of the 1981 data (as viewed in the $N-Z$ plane).  
In addition to the O and N sets, a set of 158 nuclides with more 
recently measured masses (the NB set of ``even newer'' nuclides)
is employed in the modeling process.  In earlier work [14-16,5], these 
three data sets have been used to quantify the extrapolation capability 
(the so-called extrapability) of different global mass models (based 
either on nuclear theory or neural networks).

The set ${\rm O}\, \oplus {\rm N} \, \oplus {\rm NB}$ is divided by a 
random-sampling procedure into three nonoverlapping subsets, namely 
a training set (80\%), a validation set (10\%), and a test set (10\%), 
in the indicated approximate proportions.  (In all work reported
in this paper, random samplings are drawn from a uniform
distribution.)   Training, validation, and
test sets are each further subdivided into four subsets labeled EE,
EO, OE, and OO, composed respectively of nuclides belonging to the 
four ``even-oddness'' classes: even-$Z$-even-$N$, even-$Z$-odd-$N$, 
odd-$Z$-even-$N$, and odd-$Z$-odd-$N$.  For convenience, values 
of the input variables are encoded by a linear transformation that
scales and shifts given values of $Z$ and $N$ to lie in the interval
$[0,1]$.  A similar linear transformation decodes the learning machine's 
raw output, which lies in the interval $[-1,1]$, so as to provide an 
estimate of the corresponding mass excess in MeV.

Effectively, we divide the mass problem into four separate problems,
one for each of the four ``even-oddness'' classes in $Z$ and $N$.
In doing so, we are actually incorporating some domain knowledge into 
the learning strategy.  Distinctive quantum-mechanical features of nuclei, 
abundantly supported by empirical evidence, include quantized angular 
momenta, magic numbers, shell structure, and pairing energies, 
all of which stem from the fact that $Z$ and $N$ 
are integers, even or odd.

A SVM model is developed individually for each of the four nuclear classes
EE, EO, OE, and OO.  SVM regression (with ANOVA-RBF specification of kernels) 
is carried out separately for the respective training sets, thereby
constructing a predictive model whose reliability is judged by its
performance on the examples in the test set.  Following established
practice, performance of each of the four models on its corresponding
validation set have been used to guide the final determination of the adjustable 
parameters.  Ideally, the test set should have {\it no} role in choosing
these parameters (although in some cases a weak influence is allowed).

As is usual in global models of the atomic-mass table, the quality
of a given model is judged by the smallness of the root-mean-square (rms)
error $\sigma$ in the mass excess $\Delta M$, averaged over the data 
set in question (training, validation, or test set for a given 
class of nuclides).  To be competitive, a model should have
values of $\sigma$ below 1 MeV.  It should be noted however, that
only in a few cases has a rigorous test of predictive performance
been made for the traditional theoretical models of semi-empirical
character.  (An important exception is found in the work of
M\"oller, Nix, and collaborators [15,16], who introduce the
notion of extrapability, which is equivalent to our 
generalization.)

Some of the better results obtained in the present exploratory
study are displayed in Table 1.  The performance of these models, all 
with RBF parameter $\gamma = 2.5$ and ANOVA degree $d=8$, is evidently
of high quality.  

\topinsert
\centerline{\bf{Table 1}}
\vskip 12 truept
\noindent
Performance of SVM global models of atomic mass.  For all four models,
the RBF parameter $\gamma$ is 2.5 and the ANOVA degree is $d=8$.  The
other SVM parameters have been defaulted at $C=0.1$ and $\varepsilon = 0.001$.
\vskip 27 truept

% +--------------------------------------------------------------------+
% |                                                                    |
% |                           TABLES.TEX                               |
% |                                                                    |
% |                     Ray F. Cowan  15-Feb-85                        |
% |                                                                    |
% |                       Princeton University                         |
% |                                                                    |
% |                     Last Revision: 21-Nov-85                       |
% |                                                                    |
% |   Macros I find handy for making tables.  See TABLEDOC TEX for     |
% |   a longer description.  The token-counting macros are straight    |
% |   from the TeXbook's "Dirty Tricks" appendix.                      |
% |                                                                    |
% +--------------------------------------------------------------------+
%
\newbox\hdbox%
\newcount\hdrows%
\newcount\multispancount%
\newcount\ncase%
\newcount\ncols% This is the number of primary text columns in the table.
\newcount\nrows%
\newcount\nspan%
\newcount\ntemp%
\newdimen\hdsize%
\newdimen\newhdsize%
\newdimen\parasize%
\newdimen\spreadwidth%
\newdimen\thicksize%
\newdimen\thinsize%
\newdimen\tablewidth%
\newif\ifcentertables%
\newif\ifendsize%
\newif\iffirstrow%
\newif\iftableinfo%
\newtoks\dbt%
\newtoks\hdtks%
\newtoks\savetks%
\newtoks\tableLETtokens%
\newtoks\tabletokens%
\newtoks\widthspec%
%
%  Book-keeping stuff--see how often these macros are called.
%
\immediate\write15{%
CP SMSG GJMSINK TEXTABLE --> TABLE MACROS V. 851121 JOB = \jobname%
}%
%
%  Turn on table diagnostics.
%
\tableinfotrue%
\catcode`\@=11%  Allows use of "@" in macro names, like PLAIN.TEX does.
%  Debugging aid.  Writes #1 on the
%                                    user's terminal and in the log file.
%
%  Define the \tstrut height, depth in terms of the x_height parameter.
%
\def\tstrut{\vrule height3.1ex depth1.2ex width0pt}%
\def\and{\char`\&}%  Allows us to get an `&' in the text.  This is the
%                    same as using the PLAIN TeX macro \&.
\def\tablerule{\noalign{\hrule height\thinsize depth0pt}}%
\thicksize=1.5pt%  Default thickness for fat rules.  The user should feel
%                  free to change this to his preference.
\thinsize=0.6pt%   Default thickness for thin rules.
\def\thickrule{\noalign{\hrule height\thicksize depth0pt}}%
\def\ctr#1{\hfil\ #1\hfil}%
%
%
%
%  Here are things for controlling the width of the finished table.
%
\tablewidth=-\maxdimen%
\spreadwidth=-\maxdimen%
\def\tabskipglue{0pt plus 1fil minus 1fil}%
%
%  Stuff for centering or not.
%
\centertablestrue%
%
%
%
%  \vctr vertically centers its argument in the row.
%
\parasize=4in%
\gdef\ARGS{########}%  Produces the correct number of #'s in the preamble
%                      by the time eveything is expanded and \halign sees
%                      it.
\gdef\headerARGS{####}%  Same as \ARGS, but used in \header macros.
\def\@mpersand{&}%  Allows us to get alignment tab characters later
%                   when we have made the character "&" an active macro.
{\catcode`\|=13%  Make |'s locally active.
\gdef\letbarzero{\let|0}%  Globally define a macro that allows us to
%                          keep active |'s from being expanded in edef's.
\gdef\letbartab{\def|{&&}}%
\gdef\letvbbar{\let\vb|}%
%  This \def will cause active |'s read by
%                            \ruledtable to be converted into double
%                            alignment tabs.
}%  End of locally active |'s.
{\catcode`\&=4%  Make these alignment tabs.
\def\ampskip{&\omit\hfil&}%  This local macro skips a vertical rule.
\catcode`\&=13%  Now make &'s into active macros.
\let&0%  This allows us to expand \ampskip in the next \xdef without
%        attempting to expand the & and getting an "undefined control
%        sequence" error.
\xdef\letampskip{\def&{\ampskip}}%
\gdef\letnovbamp{\let\novb&\let\tab&}
%  This will cause active &'s read by
%                                   \ruledtable to be converted into
%                                   double tabs and an \omit'ted \vrule.
}%  End of locally active &'s.
\def\begintable{%  Here we make |'s and &'s active characters so we can
%                  interpret them as macros.  Note that this action is
%                  true only until we encounter the matching \endgroup
%                  token later at the end of the \ruledtable macro.
   \begingroup%
   \catcode`\|=13\letbartab\letvbbar%
   \catcode`\&=13\letampskip\letnovbamp%
   \def\multispan##1{%  We must redefine \multispan to count the number
%                       of primary columns, not physical columns.
      \omit \mscount##1%
      \multiply\mscount\tw@\advance\mscount\m@ne%
      \loop\ifnum\mscount>\@ne \sp@n\repeat%
   }%  End of \multispan macro.
   \def\|{%
      &\omit\widevline&%
   }%
   \ruledtable%  Now we call \ruledtable to do the real work.
}%  End of \begintable macro.
\long\def\ruledtable#1\endtable{%
%
%  This macro reads in the user's data entries
%  and converts them into a ruled table.
%
%  Important note:  Many macros and parameters are re-defined here, and
%  these must be kept local to the table macros to avoid conflict with
%  their use outside of tables.  This is done by the \begingroup token
%  macro \begintable and the \endgroup token at the end of
%  this macro.
%
   \offinterlineskip%  Needed to make rules touch each other.
   \tabskip 0pt%  Needed for same reason as \offinterlineskip.
   \def\widevline{\vrule width\thicksize}%  Make outer \vrule's wider.
   \def\endrow{\@mpersand\omit\hfil\crnorm\@mpersand}%
   \def\crthick{\@mpersand\crnorm\thickrule\@mpersand}%
   \def\crnorule{\@mpersand\crnorm\@mpersand}%
   \let\nr=\crnorule%  A shorter abbreviation.
   \def\endtable{\@mpersand\crnorm\thickrule}%
   \let\crnorm=\cr%  Allows us to use \cr for our own purposes.
%
%  Cause user-typed \cr's to follow a row with a \tablerule.
%
   \edef\cr{\@mpersand\crnorm\tablerule\@mpersand}%
   \the\tableLETtokens%  Get the user's extra \let's, if any.
%
%  Put the data entries into a token register so we can scan through them
%  and see what the user is asking us to do.
%
   \tabletokens={&#1}%  We add an extra alignment tab to the beginning
%                       of the first row to allow for the first \vrule.
%
%  Now count how many rows are in the table and return the result in
%  count register \nrows; do the same for columns, and return that
%  in register \ncols.
%
   \countROWS\tabletokens\into\nrows%
   \countCOLS\tabletokens\into\ncols%
%
%  Now do a little arithmetic to convert the number of primary columns
%  into the number of physical columns that the alignment preamble must
%  prepare for;  similarly for rows.
%
   \advance\ncols by -1%
   \divide\ncols by 2%
   \advance\nrows by 1%
%
%  Tell the user how many rows and columns we found in his data, if he
%  wants to know.
%
   \iftableinfo %
      \immediate\write16{[Nrows=\the\nrows, Ncols=\the\ncols]}%
   \fi%
%
%  Now we actually go ahead and produce the table.
%
   \ifcentertables
      \ifhmode \par\fi%  Make sure we are in vertical mode.
      \line{%  The final table comes out as an \hbox of width the \hsize.
      \hss%  The final table will be centered left-to-right.
   \else %
      \hbox{%
   \fi
      \vbox{%
         \makePREAMBLE{\the\ncols}%  Generate the preamble.
         \edef\next{\preamble}%  This line and the next line force the
         \let\preamble=\next%    expansion of all \ARGS tokens into the
%                                appropriate number of #'s.
         \makeTABLE{\preamble}{\tabletokens}%  Go do the \halign here.
      }%  End of \vbox.
      \ifcentertables \hss}\else }\fi%  Finish the centering effect.
%                                       It is important that no spaces
%                                       follow the two `}' here.
%  }%  End of \line.
   \endgroup%  Return all local macros and parameters to their outside
%              values.
   \tablewidth=-\maxdimen%  Reset \tablewidth to normal.
   \spreadwidth=-\maxdimen% Same for \spreadwidth.
}%  End of macro \ruledtable.
\def\makeTABLE#1#2{%  Does an \halign for the \ruledtable macro.
   {%  Start of local parameter values.
   \let\ifmath0%     These macros would cause trouble if they were to be
   \let\header0%     expanded in the following \xdef; we \let them be
   \let\multispan0%  equal to a digit, because digits can't be expanded.
%
%  Set up the width specification here.
%
   \ncase=0%
   \ifdim\tablewidth>-\maxdimen \ncase=1\fi%
   \ifdim\spreadwidth>-\maxdimen \ncase=2\fi%
   \relax%  This \relax is absolutely necessary, without it the following
%           \ifcase will always take \ncase=0.
%
   \ifcase\ncase %
      \widthspec={}%
   \or %
      \widthspec=\expandafter{\expandafter t\expandafter o%
                 \the\tablewidth}%
   \else %
      \widthspec=\expandafter{\expandafter s\expandafter p\expandafter r%
                 \expandafter e\expandafter a\expandafter d%
                 \the\spreadwidth}%
   \fi %
%\out{Widthspec=[\the\widthspec]}%
%\out{Preamble=[\preamble]}%
   \xdef\next{%  We must force the preamble to be expanded BEFORE the
      \halign\the\widthspec{%
%        \halign is done;  this \edef\next{...}\next construction
%                does the trick.
      #1%  This is the preamble text.
      \noalign{\hrule height\thicksize depth0pt}%  Makes the top \hrule.
      \the#2\endtable%  This is the main body.
%
%     \noalign{\hrule height0.7pt depth0pt}%  Makes the last \hrule.
      }%  End of \halign.
   }%  End of \next.
   }%  End of local values.
   \next%  This \next must be outside of the local values, because now
%          we want those troublesome macros in the \let's above to have
%          their normal actions.
}%  End of macro \makeTABLE.
\def\makePREAMBLE#1{%  This macro generates the necessary preamble for a
%                      ruled table with #1 primary columns.
%                      (Primary columns means the number of columns NOT
%                       counting those used for vertical rules.)
   \ncols=#1%  Get the number of columns desired.
   \begingroup%  Start local parameter definitions.
   \let\ARGS=0%  This is the key to the whole thing; it prevents \ARGS
%                from being expanded in the following \edef's.
   \edef\xtp{\widevline\ARGS\tabskip\tabskipglue%
   &\ctr{\ARGS}\tstrut}%  A 1-column preamble.  Gets the sizing right.
   \advance\ncols by -1%  One column has been generated; decrement the
%                         counter.
   \loop%  Append as many further columns as needed to the preamble.
      \ifnum\ncols>0 %
      \advance\ncols by -1%
      \edef\xtp{\xtp&\vrule width\thinsize\ARGS&\ctr{\ARGS}}%
   \repeat
   \xdef\preamble{\xtp&\widevline\ARGS\tabskip0pt%
   \crnorm}%  Adds the last \vrule.
   \endgroup%  End of local parameters.
}%  End of macro \makePREAMBLE.
\def\countROWS#1\into#2{%  This counts the number of rows in #1 by
%                          looking for control sequences that end a row,
%                          e.g., \cr, \crthick, etc., and puts the result
%                          into count register #2.
   \let\countREGISTER=#2%
   \countREGISTER=0%
%  \out{In countROWS:  tokens are [\the#1]}%
   \expandafter\ROWcount\the#1\endcount%
}%
\def\ROWcount{%
   \afterassignment\subROWcount\let\next= %
}%
\def\subROWcount{%
%  \out{In subROWcount:  next is [\meaning\next]}%  Debugging aid.
   \ifx\next\endcount %
      \let\next=\relax%
   \else%
      \ncase=0%
      \ifx\next\cr %
         \global\advance\countREGISTER by 1%
         \ncase=0%
      \fi%
      \ifx\next\endrow %
         \global\advance\countREGISTER by 1%
         \ncase=0%
      \fi%
      \ifx\next\crthick %
         \global\advance\countREGISTER by 1%
         \ncase=0%
      \fi%
      \ifx\next\crnorule %
         \global\advance\countREGISTER by 1%
         \ncase=0%
      \fi%
      \ifx\next\header %
%     \out{In subROWcount:  next=header, ncase set=1}%
         \ncase=1%
      \fi%
%     \out{In subROWcount:  ncase is [\the\ncase]}%
      \relax%
      \ifcase\ncase %
         \let\next\ROWcount%
%        \out{subROWcount---> ncase=\the\ncase}%
      \or %
         \let\next\argROWskip%
%        \out{subROWcount---> ncase=\the\ncase}%
      \else %
      \fi%
   \fi%
%  \out{subROWcount---> NEXT=\meaning\next}%
   \next%
}%  End of macro \subROWcount.
\def\counthdROWS#1\into#2{%
\dvr{10}%
   \let\countREGISTER=#2%
   \countREGISTER=0%
\dvr{11}%
%  \out{In counthdROWS:  tokens are [\the#1]}%
\dvr{13}%
   \expandafter\hdROWcount\the#1\endcount%
\dvr{12}%
}%
\def\hdROWcount{%
   \afterassignment\subhdROWcount\let\next= %
}%
\def\subhdROWcount{%
%\out{In subhdROWcount:  next is [\meaning\next]}%
   \ifx\next\endcount %
      \let\next=\relax%
   \else%
      \ncase=0%
      \ifx\next\cr %
         \global\advance\countREGISTER by 1%
         \ncase=0%
      \fi%
      \ifx\next\endrow %
         \global\advance\countREGISTER by 1%
         \ncase=0%
      \fi%
      \ifx\next\crthick %
         \global\advance\countREGISTER by 1%
         \ncase=0%
      \fi%
      \ifx\next\crnorule %
         \global\advance\countREGISTER by 1%
         \ncase=0%
      \fi%
      \ifx\next\header %
%\out{In subhdROWcount:  next=header, ncase set=1}%
         \ncase=1%
      \fi%
%\out{In subhdROWcount:  ncase is [\the\ncase]}%
\relax%
      \ifcase\ncase %
         \let\next\hdROWcount%
%\out{subhdROWcount---> ncase=\the\ncase}%
      \or%
         \let\next\arghdROWskip%
%\out{subhdROWcount---> ncase=\the\ncase}%
      \else %
      \fi%
   \fi%
%\out{subhdROWcount---> NEXT=\meaning\next}%
   \next%
}%
{\catcode`\|=13\letbartab
\gdef\countCOLS#1\into#2{%
%  \out{In countCOLS:  tokens are [\the#1]}
   \let\countREGISTER=#2%
   \global\countREGISTER=0%
   \global\multispancount=0%
   \global\firstrowtrue
   \expandafter\COLcount\the#1\endcount%
   \global\advance\countREGISTER by 3%
   \global\advance\countREGISTER by -\multispancount
%  \out{countCOLS-->[\the\countREGISTER]}
}%
\gdef\COLcount{%
   \afterassignment\subCOLcount\let\next= %
}%
{\catcode`\&=13%
\gdef\subCOLcount{%
%\out{In subCOLcount: next is [\meaning\next]}
   \ifx\next\endcount %
      \let\next=\relax%
   \else%
      \ncase=0%
      \iffirstrow
         \ifx\next& %
            \global\advance\countREGISTER by 2%
            \ncase=0%
         \fi%
         \ifx\next\span %
            \global\advance\countREGISTER by 1%
            \ncase=0%
         \fi%
         \ifx\next| %
            \global\advance\countREGISTER by 2%
            \ncase=0%
         \fi
         \ifx\next\|
            \global\advance\countREGISTER by 2%
            \ncase=0%
         \fi
         \ifx\next\multispan
            \ncase=1%
            \global\advance\multispancount by 1%
         \fi
         \ifx\next\header
            \ncase=2%
         \fi
         \ifx\next\cr       \global\firstrowfalse \fi
         \ifx\next\endrow   \global\firstrowfalse \fi
         \ifx\next\crthick  \global\firstrowfalse \fi
         \ifx\next\crnorule \global\firstrowfalse \fi
      \fi%  End of \iffirstrow.
\relax%\out{subCOL-->  ncase=[\the\ncase]}
% \out{subCOL-->  next=\meaning\next}
      \ifcase\ncase %
         \let\next\COLcount%
      \or %
         \let\next\spancount%
      \or %
         \let\next\argCOLskip%
      \else %
      \fi %
   \fi%
%  \out{subCOL-->  countREGISTER=[\the\countREGISTER]}
   \next%
}%
\gdef\argROWskip#1{%
%  Deletes the next balanced, undelimited argument from a
%                 token list.
% \out{---> Entering argROWskip <---}
% \out{In argROWskip:  deleted arg is [#1]}%
   \let\next\ROWcount \next%
}%  End of macro \argskip.
\gdef\arghdROWskip#1{%
%  Deletes the next balanced, undelimited argument from a
%                 token list.
% \out{---> Entering arghdROWskip <---}
% \out{In arghdROWskip:  deleted arg is [#1]}%
   \let\next\ROWcount \next%
}%  End of macro \arghdROWskip.
\gdef\argCOLskip#1{%
%  Deletes the next balanced, undelimited argument from a
%                 token list.
% \out{---> Entering argCOLskip <---}
% \out{In argCOLskip:  deleted arg is [#1]}%
   \let\next\COLcount \next%
}%  End of macro \argskip.
}%  End of active &'s.
}%  End of active |'s.
\def\spancount#1{%\out{spancount--->\meaning#1}
   \nspan=#1\multiply\nspan by 2\advance\nspan by -1%
   \global\advance \countREGISTER by \nspan
%  \out{number spancount--->\the\nspan; \the\countREGISTER}
   \let\next\COLcount \next}%
\def\dvr#1{\relax}%
% \omit\hfil%
% \parindent=0pt\hsize=1.1in\valign{%
% \vfil#\vfil&\vfil#\vfil\cr\hfil\hbox{\ Added to\ }\hfil&%
% \hfil\hbox{\ empty events\ }\hfil\cr}\hfil%
\def\header#1{%
\dvr{1}{\let\cr=\@mpersand%
\hdtks={#1}%
%\out{In header:  hdtks=[\the\hdtks]}%
\counthdROWS\hdtks\into\hdrows%
\advance\hdrows by 1%
\ifnum\hdrows=0 \hdrows=1 \fi%
%\out{In header:  Nhdrows=[\the\hdrows]}%
\dvr{5}\makehdPREAMBLE{\the\hdrows}%
%\out{In header:  headerpreamble=[\headerpreamble]}%
\dvr{6}\getHDdimen{#1}%
%\out{In header:  hdsize=[\the\hdsize]}%
%\striplastCR{#1}%
{\parindent=0pt\hsize=\hdsize{\let\ifmath0%
\xdef\next{\valign{\headerpreamble #1\crnorm}}}\dvr{7}\next\dvr{8}%
}%
}\dvr{2}}%  End of macro \header.
\def\makehdPREAMBLE#1{%This macro generates the necessary preamble for a
\dvr{3}%
%                      ruled table with \ncols primary columns.
%                      (Primary columns means the number of columns NOT
%                       counting those used for vertical rules.
\hdrows=#1%  Get the number of columns desired.
{%  Start local parameter definitions.
\let\headerARGS=0%
%  This is the key to the whole thing; it prevents \ARGS
\let\cr=\crnorm%
%                from being expanded in the followin \edef's.
\edef\xtp{\vfil\hfil\hbox{\headerARGS}\hfil\vfil}%
\advance\hdrows by -1%  One row has been generated; decrement the
%                         counter.
\loop%  Append as many further rows as needed to the preamble.
\ifnum\hdrows>0%
\advance\hdrows by -1%
\edef\xtp{\xtp&\vfil\hfil\hbox{\headerARGS}\hfil\vfil}%
\repeat%
\xdef\headerpreamble{\xtp\crcr}%
}%  End of local parameters.
\dvr{4}}%  End of \makehdPREAMBLE.
\def\getHDdimen#1{%
%\out{In getHDdimen:  Arg 1=[#1]}%
\hdsize=0pt%
\getsize#1\cr\end\cr%
}%  End of macro getHDdimen.
\def\getsize#1\cr{%
%\out{In getsize:  Arg 1=[#1]}%
%  Here we have to check arg#1 and see if the first token in #1 is an
%    \end; if so, we stop, else we check the width of arg#1.
%  We recall that each arg#1 will be terminated with a \cr token.
\endsizefalse\savetks={#1}%
%\out{In getsize:  the savetks = [\the\savetks]}%
\expandafter\lookend\the\savetks\cr%
%\out{In getsize:  ifendsize = [\meaning\ifendsize]}%
\relax \ifendsize \let\next\relax \else%
\setbox\hdbox=\hbox{#1}\newhdsize=1.0\wd\hdbox%
\ifdim\newhdsize>\hdsize \hdsize=\newhdsize \fi%
%\out{In getsize:  hdsize=[\the\hdsize]}%
%\out{In getsize:  newhdsize=[\the\newhdsize]}%
\let\next\getsize \fi%
\next%
}%
\def\lookend{\afterassignment\sublookend\let\looknext= }%
\def\sublookend{\relax%
%\out{In sublookend:  looknext = [\looknext]}%
\ifx\looknext\cr %
%\out{In sublooknext:  looknext=cr}%
\let\looknext\relax \else %
%\out{In sublooknext:  looknext/=cr}%
   \relax
   \ifx\looknext\end \global\endsizetrue \fi%
   \let\looknext=\lookend%
    \fi \looknext%
}%
%
%  Allow the user to make his own names for crthick, etc.
%
\def\tablelet#1{%
   \tableLETtokens=\expandafter{\the\tableLETtokens #1}%
}%
\catcode`\@=12%  Change @'s back to their normal category code.

\nrows= 6
\ncols= 7
\begintable
{} | & \quad Learning & Set \quad \quad \quad | &  \quad Validation &  Set 
\quad \quad \quad | &  \quad \quad \quad Test &  Set \quad \quad \cr
Classes | & \# Nuclides &  $\sigma$(MeV) | & \# Nuclides  & $\sigma$(MeV) |
& \# Nuclides  & $\sigma$(MeV) \cr
EE | & 381 & 0.58 | & 48 & 0.71 | & 48 & 0.99 \cr
EO | & 360 & 0.89 | & 45 & 0.68 | & 45 & 0.62 \cr
OE | & 371 & 0.70 | & 46 & 0.78 | & 46 & 0.88 \cr
OO | & 353 & 0.75 | & 44 & 0.74 | & 45 & 0.97 
\endtable
\vskip 14 truept
\endinsert

Similar learning experiments can be found among the studies
of Ref.~[5] based on multilayer perceptrons and modified 
backpropagation training, although procedural differences
preclude direct comparisons of performance.  The best model obtained 
using O as the training set, NB as validation set, and N
as test set gave rms error figures on these sets of 0.71 MeV, 
2.28 MeV, and 2.16 MeV, respectively.  Another strategy yielded 
better results.  The set ${\rm O}\, \oplus \, {\rm N}$ was first ``purified'' by 
removing 20 nuclides with poorly measured masses.  A random sample 
M1 consisting of 1303 of the remaining 1654 examples (some 79\%) was 
used as the training set.  The complementary set, M2, played the 
role of validation set, and the NB set was used for testing
the trained model.  The best model found in this way produced
rms errors on the three sets of 0.44 MeV (M1), 0.44 Mev (M2),
and 0.95 MeV (NB).  It should be noted that this level of
performance on the mass problem was achieved after more
than a decade of successive improvements in the choices of
architectures, coding schemes, and training algorithms.

In addition to the four class-specific models SVM-EE, SVM-EO, SVM-OE,
and SVM-OO reported on in Table 1, we also constructed a single SVM 
model (denoted SVM-S) using the full O data set as the training sample, 
without making a distinction between EE, EO, OE, and OO nuclides.  
In this case, the NB nuclei are used as a validation set, guiding
the determination of the RBF and ANOVA parameters.  The parameters 
associated with the SVM-S model are again $\gamma = 2.5$ 
and $d = 8$, along with $C= 0.1$ and $\varepsilon = 0.001$.  This
model yields rms errors of 0.70 MeV on the training set O and
0.75 MeV on the validation set NB, with a $\sigma$ value of 
1.41 MeV on the N nuclei, regarded as a test set.  (These results
are erroneously cited in Ref.~[5].)  A proper averaging over
the four nuclidic classes permits a comparison between the
SVM-S model and the four models represented in Table 1.   The
composite performance of the latter models is then reflected
in $\sigma$ values of 0.73 MeV, 0.73 MeV, and 0.88 MeV
in training, validation, and testing, respectively.  

In some cases, meaningful comparisons may be drawn between the 
performance of statistical mass models based on multilayer perceptrons 
and support vector machines, and the traditional mass models based on 
nuclear theory and phenomenology.  Starting with the simple liquid-drop
model, such traditional theory-thick models have evolved over 
seven decades to achieve a high degree of sophistication and
precision.  For example, the 1992 FRDM model of M\"oller and Nix [15] 
gives $\sigma$ values of 0.67 MeV on the O set (when fitted to
this set) and 0.74 MeV on the N set (a true measure
of predictive performance of the model).  The more enhanced
FRDM model of Ref.~[16], which is fitted to the data set
${\rm M1} \, \oplus \, {\rm M2}$, yields rms errors of 0.68 MeV (M1),
0.71 MeV (M2), and 0.70 MeV (NB).  The HFB2 model of Pearson 
and collaborators [17] gives respective errors of 0.67 MeV,
0.68 MeV, and 0.73 MeV.  (We note that the result of Ref.~[17]
on the ``test set'' NB cannot be regarded as a prediction, since
the nuclei involved were used in adjusting model parameters.)

With additional refinements, it is not unreasonable to expect 
that SVM models can equal (and possibly surpass) the levels of robustness
and predictive accuracy achieved with theory-thick models and with 
multilayer perceptron models.  However, a conclusive statement
must await a thorough SVM study based on the recent AME03 mass
evaluation carried out by Audi {\it et al.}~[18]
\vskip 28 truept
\centerline{\bf 4.  SVM MODELS OF BETA-DECAY HALFLIVES}
\vskip 12 truept

\vbox{
We now turn to a second problem of regression in the statistical analysis 
of nuclear properties via support vector machines, namely fitting and
prediction of the beta-decay halflives of nuclides $(Z,N)$ that decay 100\% via 
the $\beta^-$ mode.  The data for this problem have been culled 
from the on-line repository at the Brookhaven National Nuclear Data
Center (http:$//$www.nndc.bnl.gov).  The data employed are current to May 
2005 and consist of a total of 932 examples.  Restricting 
attention to examples with halflives below $10^6$ s leaves 
633 nuclides.  When measured in seconds, the experimental values 
of $T_{1/2}$ range over 26 orders of magnitude, so it is 
more appropriate to regress $L = \log T_{1/2}$ instead of the
halflife itself, and to adopt the rms error $\sigma_L$ of the estimate
of $L$ as a figure of merit in learning, validation, and prediction
phases of the analysis.

As in the case of the mass problem, separate SVM models are
constructed for EE, EO, OE, and OO classes of nuclides.  However,
we make the simpler RBF choice of kernel, instead of pursuing
the more elaborate ANOVA option.   (Implementation based on the
ANOVA decomposition is much more demanding in terms of 
computer time.)  Each of the four data subsets (EE, EO, OE, OO) is
subdivided into training, validation, and test sets in the
approximate proportions 80\%, 10\%, and 10\%, respectively.

The results obtained from the SVM regressions are summarized in Tables 2 
and 3.  Table 2 gives the parameters and performance measures of the
models constructed for the full set of data, regardless of measured
lifetime.  Table 3 displays the corresponding results when nuclides with
$T_{1/2} \geq 10^6$ s are removed from the database.

A similar study [19] (see also Ref.~[20]) has been carried out 
with multilayer feedforward neural networks trained by ``vanilla''
backpropagation, for data available in 1995 (766 examples in total) 
However, this study did not employ the now-standard protocol in 
which a validation set is used in making the final model selection.  
Also, no subdivision into the four even-oddness classes was made.  
Instead, the full data set (or the restricted set of examples with 
$T_{1/2} < 10^6$~s) was split into a training set of approximately 
75\% of the examples and a test set consisting of the remainder.
}

\topinsert
\centerline{\bf{Table 2}}
\vskip 12 truept
\noindent
Performance of SVM global models of $\beta$-decay halflives $T_{1/2}$ 
(including examples having $T >  10^6$ s).  For all four models, 
$C=1$ and $\varepsilon =0.001$.
\vskip 30 truept

\nrows=6
\ncols=8
\begintable
         \|  \quad Learning & Set \qquad   | \quad Validation & Set \qquad  
|\qquad Test & Set \qquad | RBF kernel     \crthick
Classes  \|\# Nuclides & $\sigma_L$  |\# Nuclides & $\sigma_L$ |\# Nuclides& $\sigma_L$|$\gamma$\crthick

EE       \|  ~137      & 2.88~ | ~16        & 3.61~ |~15        & 1.72~| 5.44 \crnorule
EO       \|  ~198      & 2.75~ | ~24        & 2.27~ |~22        & 2.17~| 7.27 \crnorule 
OE       \|  ~187      & 2.37~ | ~22        & 2.76~ |~20        & 2.38~| 9.99 \crnorule
OO       \|  ~236      & 2.62~ | ~29        & 2.07~ |~26        & 2.96~| 9.55
\endtable
%\vskip 1.5truecm
\vskip 1.2truecm
\centerline{\bf{Table 3}}
\vskip 12 truept
\noindent
Performance of SVM global models of $\beta$-decay halflives (with a cutoff 
at $10^6$ s).  For all four models, $C=1$, $\varepsilon =0.001$.
\vskip 30 truept

\nrows=6
\ncols=8
\begintable
         \|  \quad Learning & Set \qquad   | \quad Validation & Set \qquad 
 | \qquad Test & Set \qquad | RBF kernel     \crthick
Classes  \|\# Nuclides & $\sigma_L$  |\# Nuclides & $\sigma_L$ |\# Nuclides& $\sigma_L$|$\gamma$\crthick

EE       \|  ~96       & 1.34~ | ~11        & 0.52~ |~10        & 1.20~| 1.78 \crnorule
EO       \|  ~140      & 0.90~ | ~17        & 0.69~ |~15        & 1.22~| 9.97 \crnorule 
OE       \|  ~122      & 1.55~ | ~14        & 0.63~ |~13        & 1.18~| 0.84 \crnorule
OO       \|  ~159      & 1.00~ | ~19        & 1.28~ |~17        & 1.34~| 8.87
\endtable
\vskip 14truept
\endinsert
\vskip 1.3truecm
%\vskip 1truecm

Comparison of the rms errors shown in Tables 2 and 3 with the
corresponding performance figures from the earlier work [19,20] shows
an improvement (reduction) in rms error values by about a factor
2, in both learning and prediction, for both the full and restricted
data sets.  Comparison may also be made with results from 
traditional nuclear theory (e.g.~Refs.~[21-23]).  Since the 
cited neural-network models could already attain performance in fitting 
and prediction comparable to that exhibited by these theory-thick models, 
we can say with some confidence that the SVM models are capable of a 
predictive acuity superior to the best of the traditional global 
models currently in play.
\vfill\eject

We should also call attention to the greatly improved quality of
neural-network models of $\beta$-decay systematics, achieved in
very recent studies [24].  Data based on the AME03 evaluation 
are divided into training, validation, and test sets in the
respective proportions 60\%, 20\%, and 20\%, both with and
without the restriction to halflives not greater than $10^6$~s,
but without subdivision into even-oddness classes.  In the
case where the restriction is imposed, the best results 
found for the error measure $\sigma_L$ are 0.55 (training),
0.61 (validation), and 0.64 (prediction).  The corresponding
averages for the model represented in Table 3 are 1.43, 0.89,
and 1.24, respectively, so further refinement of the SVM models
will be needed to match the perfomance of the best multilayer
perceptrons.
\vskip 28 truept

\centerline{\bf 5.  SVM MODELS OF GROUND-STATE SPINS AND PARITIES} 
\vskip 12 truept

In a third illustration of what is possible, the SVM approach is applied
to construct global statistical models of the ground-state spins and parities 
of nuclei.  (In this context, ``spin'' refers to the total angular momentum 
quantum number $J$ of the nuclear state.)   As in the exercises described
in Secs.~3 and 4, we again divide the nuclei under consideration into EE, 
EO, OE, and OO classes.  In the spin problem, this subdivision is of
obvious importance, since the law of angular momentum addition in
quantum mechanics dictates that the states of EE and OO nuclei can
only have integral values of $J$, whereas the spins of EO and OE 
nuclei must be half-odd-integral.  In fact, all EE nuclei are known to
have spin/parity $J^\pi = 0^+$.  Clearly, we may exclude this class from
consideration, since its modeling is a trivial task for any viable
learning machine.

The parity property of nuclear states presents the simplest kind 
of classification problem, with two mutually exclusive outcomes, even 
or odd.  Moreover, because the spin quantum number $J$ is restricted
by quantum theory to a finite set of discrete values, global modeling 
of spin systematics is also most efficiently treated, within the 
SVM framework, as a problem of classification rather than function 
approximation or regression.  In our study, we consider
$J$ values ranging from 0 to 23/2 in half-odd-integral steps, the 
integral values being available for OO nuclei and the half-odd integral 
values, for EO and OE nuclei.  This specification of the problem 
may be construed as introducing some basic domain knowledge into the 
model-building process.

Data for the spin and parity nuclear ground states have been taken from 
the on-line Brookhaven database.  Based on simple RBF kernels, separate 
SVM classifier models of these two properties have been developed for each 
of the three nontrivial even-oddness cases.

Let us first discuss our findings for the parity problem.  In treating 
this problem, the data for each of the cases EO, OE, and OO are divided 
at random into training, validation, and test sets in the approximate 
proportions 80\%, 10\%, and 10\%, respectively.  Performance is measured in
terms of the percentages of correct classifications within these
subsets.  The primary results are summarized in Table 4.  It is apparent 
that modeling parity is an easy task for SVMs.  Judging from available
results [25,14], it is also relatively easy for neural networks 
(although SVM performance is somewhat superior).

For the models of Table 4, performance on the training sets is 
perfect.  If we are willing to make a small sacrifice in the quality
of reproduction of the input data, slightly better performance on the
validation and test sets can be achieved, as seen in Table 5.
It is interesting that this second model corresponds to a quite different
error minimum under variation of the parameter $\gamma$.  In general,
there may be many such minima of similar depth.

We have not yet conducted a full training-validation-test process
for the spin problem.  Accordingly, we present only preliminary
results, which nevertheless are illuminating.  In the first experiment 
to be reported (see Table 6), each of the three spin data sets EE, OO, and OO is
divided randomly into {\it two} subsets, a training set and a 
complementary second set.  The training set contains approximately
90\% of the examples of the given even-oddness class, and the second set, the 
remaining $\sim 10$\%.

\topinsert
\centerline{\bf{Table 4}}
\vskip 12 truept
\noindent
Performance of SVM global models of ground-state parity. 
For all four models, $C=0.1$, $\varepsilon =0.01$.  Model selection
is guided by best performance on the validation set, consistent with
a perfect score on the training set.
\vskip 27 truept

\nrows=5
\ncols=6
\begintable
\|  \quad Learning & Set \qquad   | \quad Validation & Set \qquad  
 |\qquad Test & Set \qquad | RBF kernel \crthick
Classes  \|\# Nuclides & Score  |\# Nuclides & Score |\# Nuclides&
Score|$\gamma$\crthick
EO   \|  ~474     & 100\%~ | ~58        & 93\%~ |~52        & 83\%~| 9.232 \crnorule
OE   \|  ~466     & 100\%~ | ~57        & 89\%~ |~51        & 90\%~| 9.482 \crnorule
OO   \|  ~434     & 100\%~ | ~53        & 87\%~ |~48        & 84\%~| 9.176 
\endtable
\vskip 1truecm
\centerline{\bf{Table 5}}
\vskip 12 truept
\noindent
Performance of SVM global models of ground-state parity.
For all four models, $C=0.1$, $\varepsilon =0.01$.  In this case,
model selection is guided by best performance on the validation
set, allowing for minimal nonzero error rate on the training set.
\vskip 27 truept

\nrows=5
\ncols=8
\begintable
\| \quad Learning & Set \qquad   | \quad Validation & Set \qquad  | 
\qquad Test & Set \qquad | RBF kernel \crthick
Classes \|\# Nuclides & Score  |\# Nuclides & Score |\# Nuclides& Score|$\gamma$\crthick
EO  \|  ~474      & 100\%~ | ~58        & 91\%~ |~52        & 83\%~| 0.678 \crnorule
OE  \|  ~466      & 95\%~  | ~57        & 84\%~ |~51        & 92\%~| 0.180 \crnorule
OO  \|  ~434      & 96\%~  | ~53        & 83\%~ |~48        & 86\%~| 0.240
\endtable
\vskip 14truept
\endinsert
\vskip 1truecm

\topinsert
\centerline{\bf{Table 6}}
\vskip 12 truept
\noindent
Performance of SVM global models of nuclear ground-state spin. 
For all three models, $C=0.1$, $\varepsilon =0.01$.  Model selection
is guided by best on performance on the validation set, consistent with a
perfect score on the training set.
\vskip 27 truept

\nrows=5
\ncols=6
\begintable
\| \quad \quad Learning & Set  \qquad  | \quad Validation/Test & Set \qquad \
| RBF kernel  \crthick
Classes  \|\# Nuclides & Score  |\# Nuclides & Score | $\gamma$    \crthick
EO       \|  ~528      & 100\%~ | ~58        & 81\%~ | 9.217       \crnorule
OE       \|  ~522      & 100\%~ | ~57        & 68\%~ | 9.001       \crnorule
OO       \|  ~488      & 100\%~ | ~54        & 43\%~ | 4.002

\endtable
\vskip 14truept
\endinsert

\noindent
The second set is used to help pin down the RBF parameter
$\gamma$ and thereby plays a role in model selection.  Hence it must be
interpreted as a validation set.  SVM models are constructed for a range 
of $\gamma$ values, and the model whose $\gamma$ value produces the
lowest error on the second data set (while scoring 100\% on the
training set) is selected.  There is no real test set in this experiment. 
%%%%%%%%%%%%%%%%%%%%%%%%%%%%%%%%%%%%%%%%%%%%%%%%%%%%%%%%%%
\topinsert
\centerline{\bf{Table 7}}
\vskip 12 truept
\noindent
Performance of SVM global models of nuclear ground-state spin. 
For all three models, $C=0.1$, $\varepsilon =0.01$.  The
parameter $\gamma$ is fixed at the value determined for Table 6.
The test set influences model choice only indirectly.
\vskip 27 truept

\nrows=5
\ncols=6
\begintable
\|   Learning & Set    | Validation & Set   |Test & Set | RBF kernel \crthick
Classes \|\# Nuclides & Score |\# Nuclides & Score |\# Nuclides& Score~|$\gamma$\crthick
EO   \|  ~476      & 100\%~ | ~58        & 79\%~ |~52        & 60\%~~| 9.217 \crnorule
OE   \|  ~470      & 100\%~ | ~57        & 61\%~ |~52        & 79\%~~| 9.001 \crnorule
OO       \|  ~440      & 100\%~ | ~54        & 39\%~ |~48        & 38\%~~| 4.002
\endtable
\vskip 14truept
\endinsert

In an alternative experiment, we have implemented a protocol intermediate
between the training-validation scheme leading to Table 6, and the full 
training-validation-test procedure.  The data for each of the three 
even-oddness classes involved are divided into three subsets as follows.  
The second subset is taken to be identical to the second subset formed 
in the first experiment.  The first subset, used as the training set, 
consists of 80\% of the examples for the class in question, these being 
chosen at random from the corresponding training set created in 
the first experiment.  The 10\% that are
not so chosen constitute the third subset, which is regarded as
a test set.  Then, using the {\it same} parameter $\gamma$ as
determined in the first experiment with the aid of the second
subset, new SVM models are developed from the examples in the reduced 
training set.  These models are used to generate spin values for both 
second and third subsets -- values which may differ from those given by the
models developed in the first experiment (see Table 7).  Although 
it is not legitimate to interpret the third subset as a test set in the 
purest sense, its influence on model selection is indirect.

From the results shown in Tables 6 and 7, one may plausibly infer
that support vector machines can perform very well on the
problem of predicting nuclear ground-state spins.  While further
experiments are needed to affirm this conclusion, it is already
of interest to compare our SVM models with other global models
of nuclear spin systematics.  Global nuclear structure calculations
within the macroscopic/microscopic approach [26] reproduce
the ground-state spins of odd-$A$ nuclei with an accuracy of
60\% (agreement being found in 428 examples out of 713).  
(In this work, there is no clear distinction between fitting 
and prediction, or between training, validation, and test 
sets.) Multilayer feedforward neural networks do somewhat 
better [25,14].  Averaging over results of three experiments involving 
nets having a single hidden layer and trained with backpropagation, 
the performance for odd-$A$ nuclei reaches 62\% on what are 
effectively validation sets, the training sets being 
reproduced to an accuracy of 93\%.  In an experiment in which 
the connection weights of feedforward nets with one hidden layer 
are determined by a conjugate gradient procedure, performance
at the level of 99.5\% on the training set and 73.2\% on 
a validation set has been achieved for OE nuclei.  The
spins of odd-odd nuclei are notoriously difficult to predict.
This is reflected in the performance figures of neural-network
(perceptron) models on the OO category, which are typically 
75\% correct on training-set examples and only 15\% in validation or 
testing.

Placed in the context of earlier work, both statistical
and phenomenological, the results in Tables 6--7 
for the first SVM models of nuclear spin speak for themselves.
\vskip 28truept

\centerline{\bf 6.  CONCLUDING REMARKS}
\vskip 12truept

We have made initial studies of the potential of support vector machines (SVM)
for providing statistical models of nuclear systematics with demonstrable
predictive power.  Using SVM regression and classification procedures,
we have created global models of atomic masses, beta-decay halflives, 
and ground-state spins and parities.  These models exhibit performance
in both data-fitting and prediction that is comparable to that of
the best global models from nuclear phenomenology and microscopic theory,
as well as the best statistical models based on multilayer feedforward
neural networks.  Further work to develop the scope, acuity, and reliability
of SVM applications to nuclear physics seems to be warranted.  In particular,
the full body of data in the AME03 atomic-mass evaluation [18] must be brought to
bear in construction of SVM models of mass systematics, and the treatment
of the spin problem begun here needs to be completed.  Fruitful applications
to nucleon separation energies, $\alpha$-decay halflives, 
branching ratios of nuclear decay, nuclear deformations, neutron
cross sections, and other nuclear properties may also be on the horizon.
\vskip 28truept

\centerline{\bf ACKNOWLEDGMENTS}
\vskip 12 truept
This research was supported in part by the U.~S.~National 
Science Foundation under Grant No.~PHY-0140316.  For the regression
problems, we made use of the on-line mySVM software and instruction manual
of Stefan R\"uping (Dortmund) [27], and for classification problems 
we implemented the SVM-multiclass software of Thorsten Joachims 
(Cornell) [28]. 
\vskip 28truept

\centerline{\bf REFERENCES}
\vskip 12 truept
\item{[1]}
D.~E.~Rumelhart, G.~E.~Hinton, and R.~J.~Williams, in {\it
Parallel Distributed Processing: Explorations in the Microstructure
of Cognition}, Vol.~1, edited by D.~E.~Rumelhart {\it et al.} (MIT Press,
Cambridge, MA, 1986). 
\item{[2]}
J.~Hertz, A.~Krogh, and R.~G.~Palmer, {\it Introduction to the Theory
of Neural Computation} (Addison-Wesley, Redwood City, CA, 1991).
\item{[3]}
S.~Haykin, {\it Neural Networks: A Comprehensive Foundation}, Second
Edition (McMillan, New York, 1999).
\item{[4]}
J.~W.~Clark, T.~Lindenau, and M.~L.~Ristig, {\it Scientific Applications
of Neural Nets} (Springer-Verlag, Berlin, 1999).
\item{[5]}
S.~Athanassopoulos, E.~Mavrommatis, K.~A.~Gernoth, and J.~W.~Clark,
{\it Nucl.~Phys.~A} {\bf 743}, 222 (2004).
\item{[6]}
C.~Cortes and V.~Vapnik, {\it Machine Learning} {\bf 20}, 273 (1995).
\item{[7]}
V.~N.~Vapnik, {\it The Nature of Statistical Learning Theory} (Springer-Verlag, 
New York, 1995).
\item{[8]}
V.~N.~Vapnik, {\it Statistical Learning Theory} (Wiley, New York, 1998).
\item{[9]}
V.~N.~Vapnik, in {\it Advances in Neural Information Processing Systems},
Vol.~4 (Morgan Kaufmann, San Mateo, CA, 1992), p.~831.
\item{[10]}
J.~Mercer, {\it Transactions of the London Philosophical Society (A)} {\bf 209},
415 (1909).
\item{[11]}
V.~N.~Vapnik and A.~Ya.~Chervonenkis, in {\it Theoretical Probability
and Its Applications} {\bf 17}, 264 (1971).
\item{[12]}
M.~O.~Stitson, A.~Gammerman, V.~Vapnik, V.~Vovk, C.~Watkins, and J.~Weston,
in {\it Advances in Kernel Methods -- Support Vector Learning},
%#JWC:   Check Sch\"ukopf - may be Sch\"okopf.
edited by B. Sch\"ukopf, C.~Burges, and A.~J.~Smola  
(MIT Press, Cambridge, MA, 1999), p.~285.
\item{[13]}
P.~M\"oller and J.~R.~Nix, {\it At.~Data~Nucl.~Data Tables} {\bf 26}, 
165 (1981).
\item{[14]}
K.~A.~Gernoth, J.~W.~Clark, J.~S.~Prater, and H.~Bohr, {\it Phys.~Lett.} {\bf B300},
1 (1993).
\item{[15]}
P.~M\"oller and J.~R.~Nix, {\it J.~Phys.~G} {\bf 20}, 1681 (1994).
\item{[16]}
P.~M\"oller, J.~R.~Nix, W.~D.~Myers, and W.~J.~Swiatecki, 
{\it At.~Data Nucl.~Data Tables} {\bf 59}, 185 (1995).
\item{[17]}
M.~Samyn, S.~Goriely, P.-H.~Heenen, J.~M.~Pearson, and F.~Tondeur,
{\it Nucl.~Phys.} A {\bf 700}, 142 (2002);
S.~Goriely, M.~Samyn, P.-H.~Heenen, J.~M.~Pearson, and F.~Tondeur,
{\it Phys.~Rev.~C} {\bf 66}, 024326 (2002).
\item{[18]}
A.~H.~Wapstra, G.~Audi, and C.~Thibault, {\it Nucl.~Phys.~A} {\bf 729}, 337
(2003).
\item{[19]}
E.~Mavrommatis, A.~Dakos, K.~A.~Gernoth, and J.~W.~Clark, in {\it Condensed
Matter Theories}, Vol. 13, edited by J.~da Providencia and F.~B.~Malik
(Nova Science Publishers, Commack, NY, 199), p.~423.
\item{[20]}
J. W. Clark, E. Mavrommatis, S. Athanassopoulos, A. Dakos, and
K. A. Gernoth, {\it Fission Dynamics of Atomic Clusters and Nuclei}, 
edited by D.~M.~Brink, F.~F.~Karpechine, F.~B.~Malik, and J.~da Providencia 
(World Scientific, Singapore, 2001), p.~76. [nucl-th/0109081]
\item{[21]}
A.~Staudt, E.~Bender, K.~Muto, and H.~V.~Klapdor, 
{\it At.~Data Nucl.~Data Tables} {\bf 44}, 80 (1990).
\item{[22]}
H.~Homma, E.~Bender, M.~Hirsch, K.~Muto, H.~V.~Klapdor-Kleingrothaus,
{\it Phys.~Rev.~C} {\bf 54}, 2972 (1996).
\item{[23]}
P.~M\"oller, J.~R.~Nix, and K.~L.~Kratz, 
{\it At.~Data Nucl.~Data Tables} {\bf 66}, 131 (1997).
\item{[24]}
N.~Costiris, A.~Dakos, E.~Mavrommatis, K.~A.~Gernoth, and J.~W.~Clark,
to be published.
\item{[25]}
J.~W.~Clark, S.~Gazula, K.~A.~Gernoth, J.~Hasenbein, J.~S.~Prater, 
and H.~Bohr, in {\it Recent Progress in Many-Body Theories}, 
Vol.~3, edited by T.~L.~Ainsworth, C.~E.~Campbell, B.~E.~Clements, 
and E.~Krotscheck (Plenum, New York, 1992), p.~371.
\item{[26]}
P.~M\"oller and J.~R.~Nix, {\it Nucl.~Phys.~A}~{\bf 520}, 369c (1990).
\item{[27]}
S. R\"uping, mySVM, 
%${\rm http://www}$-ai.cs.uni-dortmund.de${\rm /SOFTWARE/MYSVM/}$ 
http://www-ai.cs.uni-dortmund.de/SOFTWARE/MYSVM/
(2004).
\item{[28]}
T. Joachims (2004), Multi-Class Support Vector Machine,
%${\rm http://www.cs.cornell.edu/}$ People/tj/svm\_light/svm\_multiclass.html (2004).
http://www.cs.cornell. edu/People/tj/svm\_light/svm\_multiclass.html (2004).
\bye